\documentclass[preprint,showpacs,preprintnumbers,amsmath,amssymb]{revtex4}
\usepackage{amsmath}
\usepackage{amssymb}
\usepackage{epsfig}
\usepackage{graphicx}
\begin{document}
\preprint{KEK TH-1176}
\newcommand{\beq}{\begin{eqnarray}}
\newcommand{\eeq}{\end{eqnarray}}

\newcommand{\bxsga}{B\to X_s \gamma}
\newcommand{\brbxsga}{{\cal B}(B\to X_s \gamma)}
\newcommand{\bzbzb}{ B_d^0 - \bar{B}_d^0 }

\newcommand{\bsga}{  b\to s \gamma}
\newcommand{\bdga}{  b\to d \gamma}
\newcommand{\bvga}{  B\to V \gamma }
\newcommand{\bksga}{ B\to K^* \gamma}
\newcommand{\brhoga}{B\to \rho \gamma}

\newcommand{\brbkz}{{\cal B}(B\to \overline{K}^{*0} \gamma)}
\newcommand{\brbkm}{{\cal B}(B\to K^{*-} \gamma)}
\newcommand{\brbrm}{{\cal B}(B\to \rho^- \gamma)}
\newcommand{\brbrz}{{\cal B}(B\to \rho^0 \gamma)}

\newcommand{\calb}{ {\cal B}}
\newcommand{\acp}{ {\cal A}_{CP}}
\newcommand{\oas}{ {\cal O} (\alpha_s)}

\newcommand{\mt}{m_t}
\newcommand{\mw}{M_W}
\newcommand{\mhp}{M_{H}}
\newcommand{\muw}{\mu_W}
\newcommand{\mub}{\mu_b}
\newcommand{\dmd}{\Delta M_{B_d} }
\newcommand{\ltt}{\lambda_{tt} }
\newcommand{\lbb}{\lambda_{bb} }
\newcommand{\rhob}{\bar{\rho} }
\newcommand{\etab}{\bar{\eta} }

\newcommand{\smallsm}{{\scriptscriptstyle SM}}
\newcommand{\smallyy}{{\scriptscriptstyle YY}}
\newcommand{\smallxy}{{\scriptscriptstyle XY}}
\newcommand{\smallnp}{{\scriptscriptstyle NP}}

\newcommand{\tab}[1]{Table \ref{#1}}
\newcommand{\fig}[1]{Fig.\ref{#1}}
\newcommand{\real}{{\rm Re}\,}
\newcommand{\im}{{\rm Im}\,}
\newcommand{\non}{\nonumber\\ }

\def \epjc{  Eur. Phys. J. C }
\def \jpg{  J. Phys. G}
\def \npb{  Nucl. Phys. B }
\def \plb{  Phys. Lett. B }
\def \prd{  Phys. Rev. D }
\def \prl{  Phys. Rev. Lett.  }
\def \pr{   Phys. Rep. }
\def \rmp{  Rev. Mod. Phys. }
\title{Two Higgs Bi-doublet Left-Right Model \\ With Spontaneous P and CP Violation}
\author{ Yue-Liang Wu}
\affiliation{Kavli Institute for Theoretical Physics China, Institute of Theoretical Physics, \\
Chinese Academy of Science, Beijing 100080, P.R.China}
\author{Yu-Feng Zhou}
\affiliation{KEK, Tsukuba, 305-0801, Japan}

\begin{abstract}
A left-right symmetric model with two Higgs bi-doublet is shown to
be a consistent model for both spontaneous P and CP violation. The
flavor changing neutral currents can be suppressed by the mechanism
of approximate global $U(1)$ family symmetry. We calculate the
constraints from neural $K$ meson mass difference $\Delta m_K$ and
demonstrate that a right-handed gauge boson $W_2$ contribution in
box-diagrams with mass well below 1 TeV is allowed due to a
cancellation caused by a light charged Higgs boson with a mass range
$150 \sim 300$ GeV. The $W_2$ contribution to $\epsilon_K$ can be
suppressed from appropriate choice of additional CP phases appearing
in the right-handed Cabbibo-Kobayashi-Maskawa matrix. The model is
also found to be fully consistent with $B^0$ mass difference $\Delta
m_B$, and the mixing-induced CP violation quantity $\sin2\beta_{J/\psi}$,
which is usually difficult for the model with only one Higgs
bi-doublet. The new physics beyond the standard model can be
directly searched at the colliders LHC and ILC.
\end{abstract}


\maketitle

\newpage

\section{introduction}

Since the discovery of parity (P) violation fifty years
ago\cite{LY,CSW}, it has been realized that both symmetry and
asymmetry can play important roles in particle physics. Later on,
charge-conjugation-parity (CP) violation was also discovered in kaon
decays\cite{CP}. The electroweak standard model was established
based on the left-handed symmetry $SU(2)_L$ and has well been
described by the gauge symmetry $SU(2)_L\times U(1)_Y$\cite{SM-1,SM-2,SM-3}.
Since then, one of the important issues in particle physics concerns
origin of P and CP violations as well as the smallness of flavor
changing neutral currents(FCNC). Its solution requires physics
beyond the standard model.

The investigation of explicit CP violation in the standard model led
to the prediction for the existence of three generation quarks, so
that a single Kobayashi-Maskawa CP phase\cite{KM} can be introduced
to characterize the CP-violating mechanism in the standard model.
Such a simple CP-violating mechanism has been found to be remarkable
for explaining not only the indirect CP violation in kaon decays,
but also the direct CP violation in kaon decays\cite{YLW1-1,YLW1-2}
observed by two experimental groups at CERN\cite{NA48} and
Fermilab\cite{KTeV}, and the direct CP violations in B meson
decays\cite{WZZ1,WZZ2,WZZ3} reported by two
B-factories\cite{BB1,BB2}. Nevertheless, the CP violation in the
standard model is assumed to be caused from the explicit complex
Yukawa couplings put in by hand, thus its origin remains unknown. To
understand the origin of CP violation, a
spontaneous CP violation mechanism was suggested by 
Lee in 1973\cite{TDL-1,TDL-2} in which scalar fields are responsible to CP
violation. Soon after, an interesting spontaneous CP-violating three
Higgs doublet model was proposed by imposing discrete
symmetries\cite{3HDM} in order to avoid the FCNC, while such a model
has been strongly constrained from the low energy phenomena of K and
B systems. By abandoning the natural flavor conservation
hypothesis\cite{HW,WW,WU}, a general two Higgs doublet model (2HDM)
motivated from spontaneous CP violation has been investigated in
detail\cite{WW,WU}, where the FCNC is assumed to be naturally
suppressed by the mechanism of approximate global U(1) family
symmetry\cite{WW,WU}. Of particular, it has been shown in
refs.\cite{WW,WU} that after spontaneous symmetry breaking the
single relative CP phase of two vacuum expectation values can induce
rich CP-violating sources, which not only explain the KM
CP-violating mechanism in the standard model, but also lead to new
type of CP-violating sources in the charged Higgs interactions. Such
a model can result in new physics
phenomena\cite{WW1,WU1,WZ1,WZ2,WZ3} and remains consistent with the
current experiments.

With the hypothesis that parity is a good symmetry at high energy, a
left-right symmetric model was proposed based on the gauge group
$SU(2)_L\times SU(2)_R \times U(1)_{B-L}$\cite{LRM1,LRM2,LRM3}. In
such a model, parity violation can naturally be understood via
spontaneous symmetry breaking. Also CP asymmetry can be realized as
a consequence of spontaneous symmetry
breaking\cite{LRM4,LRM5,LRM6,LRM7}. Nevertheless, the spontaneous P
and CP-violating left-right model with only one Higgs bi-doublet is
strongly constrained\cite{LRM5,LRM6,LRM7} from low energy
phenomenology:

(i) The neutral kaon mass difference $\Delta m_K$ requires that the
right-handed gauge bosons must be very heavy above 2 TeV to suppress
the extra box-diagram as the gauge coupling is left-right symmetric. Since
the Yukawa couplings for neutral and charged Higgs bosons are fixed
to quarks masses and Cabbibo-Kobayashi-Maskawa (CKM) matrices. There
is no cancellation occurring among different contributions.
For the same reason, the lightest neutral Higgs boson must be above
10 TeV\cite{LRM7,FCNC,NSCP1} to suppress FCNC. Such a neural Higgs
mass is too heavy in the Higgs bi-doublet sector to make the model
natural as the bi-doublet Higgs bosons are expected to be at the
electroweak scale which is much lower than the right-handed gauge
boson mass;
(ii) In the one Higgs bi-doublet model, all the CP violating phases
are calculable quantities in terms of quarks masses and ratios of
VEVs, which can be directly tested by the experimental data on CP
violating observables. It has been shown \cite{LRM7} that the
combing constraints from $K$ system and $B$ system actually excluded the
so-called minimal one Higgs bi-doublet left-right model with
spontaneous CP violation in the decoupling limit, as the model fails
to reproduce the precisely measured weak phase angle $\sin2\beta$
from B factories;
(iii) Furthermore, the condition for the spontaneous CP violation
requires an unnatural fine tuning of the Higgs potential in the one
Higgs bi-doublet LR model\cite{NSCP1,NSCP2}.  For those reasons, it
was motivated to consider the one Higgs bi-doublet LR model with
general CP violation\cite{LRM8,LRM9,LRM10} instead of spontaneous CP
violation. An alternative consideration for spontaneous P and CP
violation is to introduce the concept of mirror
particles\cite{MR-1,MR-2}, recently, a maximally symmetric
model\cite{SO32-1,SO32-2} was constructed along this line by
considering mirror quarks and leptons.
%
%
%

In this note, motivated by the general 2HDM as a model for
spontaneous CP violation, we shall simply extend the one Higgs
bi-doublet left-right model to a two Higgs bi-doublet left-right
model with spontaneous P and CP violation, and demonstrate that the
above mentioned stringent phenomenological constraints from neutral
meson mixings can be significantly relaxed. It will be shown that
the right-handed gauge boson mass can be as low as 600 GeV with the
charged Higgs mass around 200 GeV.  The FCNC will not impose severe
constraints on the neural Higgs mass, provided small off-diagonal
Yukawa couplings via the mechanism of approximate global $U(1)$
family symmetry\cite{HW,WW,WU}.

The paper is organized as follows:
in section II, we present a general description for two Higgs
bi-doublet left-right model. In section III, we analyze the neutral
$K$ system, which includes the mass difference $\Delta m_K$ and
indirect CP violation $\epsilon_K$. We observe that the right-handed
gauge boson contributions to the mass difference $\Delta m_K$ can be
opposite to that from the charged Higgs boson in this extended model
and a cancellation between the two contributions is possible in a
large parameter space.
The suppression  of right-handed gauge boson contributions to the
indirect CP violation $\epsilon_K$ is found to occur naturally.  As
a consequence, a light right-handed gauge boson around the current
experimental low bound is allowed. In section IV, we discuss in
detail the neutral B meson system, the mass difference $\Delta m_B$
and the time dependent CP asymmetry in $B^0\to J/\Psi K_S$ decay are
found to be consistently characterized in the two Higgs bi-doublet
model with spontaneous P and CP violation, which is unlike the one
Higgs bi-doublet model.  Conclusions and remarks are presented in
the last section.

\section{Properties of a Two Higgs bi-doublet Left-right Model }


 The left-handed and right-handed quarks and leptons in the $SU(2)_L
 \otimes SU(2)_R \otimes U(1)_{B-L}$ model are all given by the
 doublets

\begin{eqnarray}
\label{Q} Q_{iL}&=& \left(
                    \begin{array}{c}
                      u_i \\
                      d_i \\
                    \end{array}
                  \right)_L
 \;\;:\;(2,1,1/3),\;\;\;\; Q_{iR}= \left(
                    \begin{array}{c}
                      u_i \\
                      d_i \\
                    \end{array}
                  \right)_R  :\quad (1,2,1/3),
\nonumber\\
\label{L} L_{iL}&=& \left(
                    \begin{array}{c}
                      \nu_i \\
                      l_i \\
                    \end{array}
                  \right)_L : \quad (2,1,-1),\quad L_{iR}=\left(
                    \begin{array}{c}
                      \nu_i \\
                      l_i \\
                    \end{array}
                  \right)_R : \quad (1,2,-1),
\end{eqnarray}
where $i=1,2,3$ runs over number of generations. The quantum numbers
($X_L, X_R, Y$) in parenthesis denote the $SU(2)_L$, $SU(2)_R$ and
$U(1)_{B-L}$ representation. $X_{L,R}$ represent dimensions of the
$SU(2)_L$ and $SU(2)_R$ representations, and $Y$ is the hypercharge
$Y=B-L$. As a gauge invariant model, three gauge fields for the
symmetry group $SU(2)_L\times SU(2)_R\times U(1)_{B-L}$ are
introduced as $W^\mu_L$, $W^\mu_R$ and $B^\mu$ respectively.
The gauge invariant fermion-gauge interactions are constructed as
follows
\begin{equation}\label{l1}
  \mathcal{L}_f=\sum\limits_{\Psi=(Q),(L)}\bar{\Psi}_L\gamma^{\mu}
  \left( i\partial_{\mu}
    +g_L\frac{\tau^i}{2}W^i_{L\mu}+g^{\prime}\frac{Y}{2}B_{\mu} \right)
  \Psi_L + \left( L \rightarrow R \right).
\end{equation}




To generate masses of fermions and gauge bosons, we shall introduce
scalar fields and apply the Higgs mechanism to break symmetry
spontaneously. In order to generate fermion mass matrices, one only
needs to introduce one Higgs bi-doublet \cite{LRM2,LRM3}.  However,
in view of the above mentioned phenomenological difficulties, here
we shall consider a left-right symmetric model with two Higgs
bi-doublets

\begin{equation}\label{bidublet}
  \phi = \left(
    \begin{array}{cc}
      \phi_1^0  & \phi_1^+ \\
      \phi_2^- & \phi_2^0
    \end{array} \right),
  \quad
  \chi = \left(
    \begin{array}{cc}
      \chi_1^0  & \chi_1^+ \\
      \chi_2^- & \chi_2^0
    \end{array}
  \right)
  \quad :
  \left( 2,2,0 \right) .
\end{equation}
The most general Yukawa interaction for quarks is given by
\begin{eqnarray}\label{yukava}
  \mathcal{L}_Y & = & - \sum\limits_{i,j}\bar{Q}_{iL}
  \left( (y_q)_{ij}\phi+ (\tilde{y}_q)_{ij}\tilde{\phi}  +
    (h_q)_{ij}\chi+ (\tilde{h}_q)_{ij}\tilde{\chi} \right) Q_{jR} ,
\end{eqnarray}
where $\tilde{\phi}(\tilde{\chi})  =  \tau_2\phi^{\ast}(\chi^{\ast})\tau_2$
also belong to the representation $(2,2,0)$.
Parity $P$ symmetry requires $ g_L =g_R \equiv g$ and
\begin{eqnarray}
 y_{q}&=&y^\dagger_{q},\quad \tilde{y}_{q}=\tilde{y}^\dagger
_{q}, \quad h_{q}=h^\dagger_{q},\quad
\tilde{h}_{q}=\tilde{h}^\dagger_{q}.
\end{eqnarray}
When both P and CP are required to be broken
down spontaneously, all the Yukawa coupling matrices are real
symmetric.

Note that allowing the two Higgs bi-doublet coupling to the same
quark field may generate large FCNC at tree level. To suppress FCNC,
we shall follow the similar treatment in the general
two-Higgs-doublet model\cite{WW,WU} by considering the mechanism of
approximate global $U(1)$ family symmetry\cite{HW,WW,WU}
\begin{align}
(u_i, d_i)\to e^{-i\theta_i} (u_i, d_i) ,
\end{align}
which is motivated by the approximate unity of the CKM matrix. As
an consequence,  $y$, $\tilde{y}$, $h$ and $\tilde{h}$ are nearly
diagonal matrices.

To break $SU(2)_L\otimes SU(2)_R \otimes U(1)_{B-L}$ to the
$U(1)_{em}$, it requires to introduce other Higgs multiplets in
addition to  $\phi$ and $\chi$. The most
popular choice for generating small neutrino masses is to introduce
Higgs triplets $(\Delta_L$ $\sim (3,1,2)$, $\Delta_R$ $\sim (1,3,2)
)$ \cite{LRM2,LRM3}
\begin{equation}
\label{triplet} \Delta_{L,R}=
\left(
  \begin{array}{cc}
    \delta_{L,R}^+/\sqrt{2} &  \delta_{L,R}^{++} \\
    \delta_{L,R}^0 & -\delta_{L,R}^+/\sqrt{2} \\
  \end{array}
e\right) .
\end{equation}
Introducing the SU(2) triplets breaks the custodial symmetry and
leads to corrections to the parameter
$\rho=m_W^2/(m_Z^2\cos^2\theta_W)$ from unity at tree level, which
may subject to strong constraints from the LEP data. However, the
corrections from the model involve a number of free parameters such
as the ratios among the VEVs of Higgs bidoublets and triplets. The
constraint from a single $\rho$ parameter is not severe.
Furthermore, the low energy process such as $\mu$ decays and neutral
current interactions $\nu N$ , $\nu e$ and $e N$ are all affected by
the model parameters, which modifies the SM relations among the
electroweak precision observables. Useful constraints can be
obtained from a global fit to all the relevant data.  It has been
shown that the combined analysis on both the high and low energy
electroweak data within the minimal left-right model only leads to a
mild lower bound of $M_2 >700\sim 800$ GeV for right-hande gauge
boson $W_2$ when the correction to $\rho$ parameter is below $1\%$.
Since the two Higgs bidoublet model considered here contains one
more bidoublet which does not violate the custodia symmetry and
contain more parameters, the constraints should be even weaker.

The most general form of Higgs potential in this model is rather
complicated, which involves the quadratic and quartic terms for the
extra bi-doublet field $\chi$ and its mixing with $\phi$ and the two
triplets $\Delta_{L,R}$. It has been shown that by simply adding a
singlet Higgs field to the one Higgs bi-doublet LR model, the
spontaneous CP violation can occur naturally\cite{BL}. The two Higgs
bi-doublet model has definitely more flexibility in Higgs potential.
It is expected that in the most general case the spontaneous CP
violation is allowed, which will not be discussed in detail in the
present note.
%
%

After the spontaneous symmetry breaking, the two Higgs
bi-doublet fields can have the following vacuum expectation values(VEVs)
\begin{equation}
\label{bidublet1} \langle\phi \rangle = \left( \begin{array}{cc}
                          v_1e^{i\delta_1}  & 0 \\
                          0 & v_2e^{i\delta_2}
                        \end{array} \right)
                        \quad \mbox{and}\quad
                        \langle\chi \rangle = \left( \begin{array}{cc}
                          w_1 e^{i\varphi_1} & 0 \\
                          0 & w_2 e^{i\varphi_2}
                        \end{array} \right) ,
\end{equation}
which leads to  the following mass matrix for quarks
\begin{eqnarray}
M_u & = & y_q v_1 e^{i\delta_1} + \tilde{y}_q v_2
e^{-i\delta_2} + h_q w_1 e^{i\varphi_1} + \tilde{h}_q w_2
e^{-i\varphi_2},
\nonumber\\
M_d & = & y_q v_2 e^{i\delta_2} + \tilde{y}_q v_1
e^{-i\delta_1} + h_q w_2 e^{i\varphi_2} + \tilde{h}_q w_1
e^{-i\varphi_1} .
\end{eqnarray}
As the mass matrices are symmetric, they can be diagonalized by
\begin{eqnarray}
  M_u = U M^D_u U^T
  \quad\mbox{and}\quad
  M_d = V M^D_d V^T ,
\end{eqnarray}
with $M^D_{u(d)}$  being diagonal mass matrices for up(down)-type quarks.
It follows that the resultant
quark mixing matrices for left-handed and right-handed quarks are
complex conjugate to each other
\begin{eqnarray}
  K^L = U^\dagger V
  \quad\mbox{and}\quad
  K^R = U^T V^* =K_L^{\ast} .
\end{eqnarray}
Note that rotating the left-handed quark mixing matrix to the
standard CKM form $V^L$ with a single CP phase is non-trivial due to
the existence of right-handed quark-gauge interactions and charged
Higgs Yukawa interactions. In this case, there are in general five
additional CP phases $\alpha_i, (i=1,2,3)$ and $\beta_i, (i=1,2)$ in
the right-handed quark mixing matrix which is parametrized as
follows
\begin{align}
V^{R} & =\eta^{u}\left(\begin{array}{lll}
(V^{L}_{ud})^{*}e^{2i\alpha_{1}} & (V^{L}_{us})^{*}e^{i(\alpha_{1}+\alpha_{2}+\beta_{1})} & (V^{L}_{ub})^{*}e^{i(\alpha_{1}+\alpha_{3}+\beta_{1}+\beta_{2})}\\
(V^{L}_{cd})^{*}e^{i(\alpha_{1}+\alpha_{2}-\beta_{1})} & (V^{L}_{cs})^{*}e^{2i\alpha_{2}} & (V^{L}_{cb})^{*}e^{i(\alpha_{2}+\alpha_{3}+\beta_{2})}\\
(V^{L}_{td})^{*}e^{i(\alpha_{1}+\alpha_{3}-\beta_{1}-\beta_{2})}&
(V^{L}_{ts})^{*}e^{i(\alpha_{2}+\alpha_{3}-\beta_{2})} &
(V^{L}_{tb})^{*}e^{2i\alpha_{3}}\end{array}\right)\eta^{d} .
\end{align}
The sign matrices $\eta^{u,d}$ corresponds to the 32 different sign
arrangements of quark masses\cite{LRM7}. In the following
considerations, we should focus only on the case with positive quark
masses, i.e. $\eta^u=\eta^b=\openone$.

Within the Wolfenstein parametrization, we define $\beta_L \equiv
\mbox{arg}(V^{L*}_{td}V_{tb}^L )$ which is, to a high precision, one
of the angles of the unitarity triangle. As $V_{tb}^L$ is real by
convention, one has $\beta_L=\mbox{arg}(V^{L*}_{td})$. Under this
parametrization, one can define a similar quantity $\beta_R \equiv
\mbox{arg}(V^{R*}_{td} V_{tb}^R)$.  They satisfy the following
relation
\begin{eqnarray}
\beta_L + \beta_R = -(\alpha_1-\alpha_3-\beta_1-\beta_2) ,
\end{eqnarray}
which is useful for discussing $B$ meson system. Similarly, one can
define phase parameters relevant to the $K$ meson system, i.e.,
$\beta'_L \equiv \mbox{arg}(V^{L*}_{td}V_{ts}^L )$ and $\beta'_R \equiv
\mbox{arg}(V^{R*}_{td} V_{ts}^R)$. They are related by
\begin{eqnarray}
  \beta'_L + \beta'_R = -(\alpha_1-\alpha_2-\beta_1) .
\end{eqnarray}
In the Wolfenstein parametrization, it is known that
$|\mbox{arg}(V_{ts}^L )/\mbox{arg}(V^{L}_{td})| \ll 1 $, we have in a good
approximation
\begin{eqnarray}
\beta'_L \simeq \beta_L
\quad \mbox{and} \quad
\beta'_R = \beta_R +
\alpha_2-\alpha_3 - \beta_2 .
\end{eqnarray}

With enlarged Higgs sector, in this model there are two doubly charged
Higgs $H^{++}_i,(i=1,2)$, four singly charged Higgs particles $H^{+}_i,(i=1,\dots 4)$,
six neutral scalars $h^{0}_i,(i=1,\dots 6)$ and
four neural pseudo-scalars $A^{0}_i,(i=1,\dots 4)$.
The doubly charged $H^{++}_i$  contribute  only to the leptonic sector
such as lepton flavor violation processes\cite{Akeroyd:2006bb}, whereas the
singly charged and neutral scalars may have significant effects on
mixings and CP violation in quark sector.
For the sake of simplicity, we should work in a simplest scenario
that only one charged Higgs (labeled as $H^+$) is light enough to
actively contribute to the box-diagrams. Of particular, when the
VEVs satisfy the conditions $v_2\ll v_1$ and
$w_2\ll w_1$ or more precisely $v_2/v_1, w_2/w_1 < m_b/m_t$, which
are also needed for obtaining the experimentally allowed small
mixing between left-handed and right-handed gauge bosons, then many
features of this model are similar to the general 2HDM with
spontaneous CP violation\cite{WW,WU}. Therefore, we consider here
the 2HDM-like charged Higgs to be the lightest one, the
corresponding Yukawa interaction is  parametrized as follows
\begin{eqnarray}
  \mathcal{L}_{C}&=&-(2\sqrt{2}G_{F})^{1/2}\bar{u}^i
  \left(
    \sqrt{m^u_{i} m^u_{k}} \ \xi^u_{ik} V^{L}_{kj}P_{L}-V^{L}_{ik}
    \sqrt{m^d_{k} m^d_{j}}\ \xi^d_{kj} P_{R}
  \right) d^j\ H^{+} +\mbox{H.c} .
\end{eqnarray}
Here we have used the Cheng-Sher parametrization\cite{CS} in the
general 2HDM with $\xi^{u(d)}_{ij}$ the effective Yukawa coupling
matrices in the physics basis after spontaneous symmetry breaking.
The small off-diagonal terms characterized in $\sqrt{m_{i}^q
  m_{j}^q}\ \xi^{q}_{ij}$ describe the breaking of the
global $U(1)$ family symmetry. For future convenience, we denote
the diagonal elements as $\xi_c\equiv \xi^u_{22}$ and $\xi_t\equiv \xi^u_{33}$ etc.

For the flavor changing neutral Higgs boson interactions, in the
same case that $v_2/v_1, w_2/w_1 < m_b/m_t$, it becomes similar to
the general 2HDM with spontaneous CP violation. When considering the
2HDM-like neutral Higgs boson $h^0$ to be the lightest one, the
dominant interactions can approximately be expressed in the
following form
\begin{equation}
\mathcal{L}_{N}=-(\sqrt2 G_F)^{1/2}\bar{q}^{i}_L \sqrt{m_{i}^{q}
m_{j}^{q}}\ \eta_{ij}^{q}\  q_{R}^j\ h^{0}+\mbox{H.c}
\end{equation}
%
%
where $\eta^{q}_{ij}$ are given by $\xi^{q}_{ij}$ up to the factors
caused by the mixing matrix elements $O_{ij}$ among the neutral
Higgs bosons, i.e., $\eta_{ij}^{q} \sim (O_{1k} \pm i O_{1l})
\xi^{q}_{ij}$. Note that a remarkable difference from the one Higgs
bi-doublet LR model with spontaneous CP violation is that the
effective Yukawa couplings $\xi^{q}_{ij}$ or $\eta_{ij}^{q}$ in the
physics basis are in general all complex and no longer symmetric due
to spontaneous P and CP violation, they contain more free parameters
due to the extra source of CP violation in the VEVs and more Yukawa
couplings associated with the extra Higgs bi-doublet. As a
consequence, the effective Yukawa couplings $\eta_{ij}^{q}$ or
$\xi^{q}_{ij}$ and the CKM matrices $V^L$ and $V^R$ are no longer
directly linked to the quark masses and the ratios of VEVs. Similar
to the two Higgs doublet model with spontaneous CP
violation\cite{WW,WU}, the effective Yukawa couplings
$\eta_{ij}^{q}$ or $\xi^{q}_{ij}$ in the two Higgs bi-doublet model
are also free parameters which can cause significantly different
effects in low energy phenomenology.

The neutral meson mixing can arise from the neutral scalar exchange
at tree level, which could be significant. The contributions to the
mixing matrix between the neutral meson $P^{0}$ and $\bar{P}^{0}$
can easily be obtained. Denoting $P^{0}$ the bound state of two
quarks with quantum number $P^0 \equiv
(\bar{q}_{i}\gamma_{5}q_{j})$, we have, in the factorization
approximation, the following general form
\begin{equation}\label{fcnc}
M_{12}=\frac{1}{2 m_P}\langle P^0 | H_{eff} | \bar{P}^0 \rangle
\simeq G_{F}\frac{\sqrt{2}f_{P}^{2}m_{P} B^S_P}{4m_{h^0}^{2}}
\left[
\frac{1}{6}+\frac{m^q_{i}m^q_{j}}{\left(m^q_{i}+m^q_{j}\right)^{2}}
\right]
(\eta_{ij}^{q}-\eta_{ji}^{q*})^{2} .
\end{equation}
%


\section{Neutral $K$ meson mixing}

We proceed to discuss the low energy phenomenological constraints of
this model. Since $W_{L} W_{R}$ mixing angle is very small from
$\mu$ decays, the mass eigenstate $W_{1(2)}$ is almost
left(right)-handed.
In the left-right model, the strongest constraint comes from $K$
meson system.  The $K^0$ meson receives additional contributions
from both $W_1 W_2$ loop and charged Higgs loop in box-diagrams. As
the internal $(c,c)$ quark loop dominates the whole contribution,
any CP violating phases associated with it have to be strong
suppressed in order to accommodate the tiny CP violating parameter
$\epsilon_K$. In $K^0$ mixing, the $W_1 W_2$ box-diagrams are
proportional to the  following CKM factor combinations
\begin{eqnarray}
\lambda^{LR}_{q}\lambda^{RL}_{q'}=V^L_{qs}V^{R*}_{qd} V^R_{q's}V^{L*}_{q' d} \ \ \ & & (q,q'=u,c,t),
\end{eqnarray}
The condition for $(c,c)$ loop to be CP conserving leads to
\begin{eqnarray}\label{phase}
\alpha_1-\alpha_2-\beta_1 \simeq 0 \ \ \mbox{ or } \  \ \beta'_R
\simeq -\beta'_L \simeq -\beta_L .
\end{eqnarray}
As all the CP phases $\alpha_i$ and $\beta_i$ are expected to be
small quantities, we neglect the possibility of
$\alpha_1-\alpha_2-\beta_1 \simeq \pi$.
For the charged Higgs contribution, the situation is similar to the
general 2HDM: although the Yukawa couplings are
less constrained and in general complex, the dominant contribution is
only proportional to the left-handed CKM matrix $V^L$ in the same
manner as in the SM.  Thus the charged Higgs contribution to $(c,c)$
loop remains real up to $\mathcal{O}(\lambda^5)$, with $\lambda$ the
Wolfenstein parameter.

Since both of the two contributions are nearly real, their interference
is either constructive or destructive. It will be shown bellow that
due to the different chiralities, the contribution from charged
Higgs loop interferes always destructively with the $W_1 W_2$ loop
in the CP conserving case of Eq.(\ref{phase}). This provides a
possibility of a nearly  complete cancellation, which may greatly
reduce the mass lower bounds for both $W_2$ and charged Higgs $H^+$.

The SM $W_1W_1$ loop diagram contribution to $K^0-\bar{K}^0$ mixing
is described by the following effective Hamiltonian
\begin{eqnarray}
H^{W_1 W_1}_{eff} & = & \frac{G_{F}^{2}m_{W}^{2}}{16\pi^{2}}\left[(\lambda_{c}^{LL})^{2}
\eta_{cc}S_{0}(x_{c})+(\lambda_{t}^{LL})^{2}\eta_{tt}S_{0}(x_{t})
+2\lambda_{c}^{LL}\lambda_{t}^{LL}\eta_{ct}S_{0}(x_{c},x_{t})\right]
\nonumber\\
 &  & \bar{d}\gamma^{\mu}(1-\gamma_{5})s\otimes \bar{d}\gamma_{\mu}(1-\gamma_{5})s+\mbox{H.c},
\end{eqnarray}
where $m_W$ is the mass of the left-handed gauge boson and
\begin{eqnarray}
S_{0}(x) & = & \frac{x}{(1-x)^{2}}\left[1-\frac{11x}{4}+\frac{x^{2}}{4}-\frac{3x^{2}\ln x}{2(1-x)}\right] ,
\\
S_{0}(x_{c},x_{t}) & = &
x_{c}x_{t}\left[-\frac{3}{4(1-x_{c})(1-x_{t})}+\frac{\ln
x_{c}}{(x_{c}-x_{t})(1-x_{c})^{2}}\left(1-2x_{c}+\frac{x_{c}^{2}}{4}\right)+(x_{c}\leftrightarrow
x_{t})\right] . \nonumber
\end{eqnarray}
the Inami-Lim functions \cite{Inami:1981fz}.  The CKM
factors are defined as $\lambda_{q}^{LL}=V_{qs}^{L}V_{qd}^{L*}$. The
matrix element is given by
\begin{equation}
\langle K^{0}|\bar{d}\gamma^{\mu}(1-\gamma_{5})s\otimes\bar{d}\gamma_{\mu}(1-\gamma_{5})s|\bar{K}^{0}\rangle
=\frac{8}{3}f_{K}^{2}m^2_{K}B_{K} ,
\end{equation}
with the normalization $f_{K}=159.8$ MeV . The values for QCD
corrections are $\eta_{cc}\simeq1.46$, $\eta_{ct}=0.47\pm0.04$, and
$\eta_{tt}=0.5765\pm0.0065$
\cite{QCDC}, and the  bag parameter is $B_{K}=0.86\pm0.06\pm0.14$\cite{Bag}.


The effective Hamiltonian for $W_{1} W_{2}$ loop and
$S(\mbox{Goldstone}) W_2$ has been extensively investigated
\cite{LRB1,LRB2,LRB3,LRB4,LRB5,LRB6,LRB7,LRB8} and reads
\begin{eqnarray}
H^{W_1W_2+SW_2}_{eff} & = & \frac{G_{F}^{2}m_{W}^{2}}{8\pi^{2}}\beta
\sum_{i,j}\lambda_{i}^{LR}\lambda_{j}^{RL}\sqrt{x_{i}x_{j}}\left[(4+x_{i}x_{j}\beta)\eta^{LR}_1
I_{1}(x_{i},x_{j},\beta) \right.
\nonumber\\
&& \left. -(1+\beta)\eta^{LR}_2 I_{2}(x_{i},x_{j},\beta)\right]
    \bar{d}(1-\gamma_{5})s \otimes \bar{d}(1+\gamma_{5})s+\mbox{H.c},
\end{eqnarray}
where $x_{i}=m_{i}^{2}/m_{W}^{2}$ and $\beta=m_{W}^{2}/M_{2}^{2}$.
The two QCD correction coefficients are $\eta^{LR}_1=1.4$ and
$\eta^{LR}_2=1.17$ for $\Lambda_{QCD}=0.2$GeV\cite{LRM5}.
The phases in CKM matrix elements are defined as
$\lambda_{q}^{LR}=V_{qs}^{L}\left(V_{qd}^{R}\right)^{*}$ and
$\lambda_{q}^{RL}=V_{qs}^{R}\left(V_{qd}^{L}\right)^{*}$
with the  loop functions\cite{LRB7}
\begin{eqnarray}
I_{1}(x_{i},x_{j},\beta) & = & \frac{x_{i}\ln x_{i}}{(1-x_{i})(1-x_{i}\beta)(x_{i}-x_{j})}+(i\to j)
-\frac{\beta\ln\beta}{(1-\beta)(1-x_{i}\beta)(1-x_{j}\beta)}
\nonumber\\
I_{2}(x_{i},x_{j},\beta) & = & \frac{x_{i}^{2}\ln
x_{i}}{(1-x_{i})(1-x_{i}\beta)(x_{i}-x_{j})}+(i\to
j)-\frac{\ln\beta}{(1-\beta)(1-x_{i}\beta)(1-x_{j}\beta)} .
\end{eqnarray}
In the limit of $x_i=x_j$ and $\beta\ll1$, they reduce to
\begin{eqnarray}
I_{1}(x,\beta) & \simeq & \frac{1}{(1-x)}+\frac{\ln x}{(1-x)^{2}}-\beta\ln\beta ,
\\
I_{2}(x,\beta) & \simeq & \frac{x}{1-x}+\frac{(2-x)x\ln x}{(1-x)^{2}}-\ln\beta .
\end{eqnarray}

Compared with the loop functions $S_0$ for $W_1W_1$ loop, the
functions $I_1(x,\beta)$ and $I_2(x,\beta)$ have different
mass dependencies:
%
(i) For very small $x_{c}\ll1$, the loop functions can be further
simplified as $I_{1}(x_{c},\beta)\simeq\ln x_{c}+1$, and
$I_{2}(x_{c},\beta)\simeq-\ln\beta$. It is easy to see that the
combination $4\eta^{LR}_1 I_1(x_c,\beta)-\eta^{LR}_2 I_2(x_c,\beta)$
is always $negative$. The sign of the amplitude is of crucial
importance when there are multiple sources of contributions. The
negative contribution may lead to cancellations with other amplitudes
such as Higgs loops.
%
(ii) The functions grow slowly with the internal quark masses. The
typical values (for $M_{2}=1$ TeV) are $\beta
x_{c}I_{1(2)}(x_{c},\beta)=-1.3\times10^{-5}(9.2\times10^{-6})$
and $\beta x_{t}I_{1(2)}(x_{t},\beta)=-4.3\times10^{-3}(0.077)$.
Comparing with $S_0(x_c(x_t))=2.8\times 10^{-4}(2.6)$,
one sees that for $t-$quark loop, the loop
functions $I_{1}(x_{t},\beta)$ and $I_{2}(x_{t},\beta)$ are much
smaller than $S_0(x_{t})$, which significantly suppresses the
$W_{1}W_{2}$ loop contribution to $B^{0}$ mixing. While for
$c-$quark loop, the $W_{1}W_{2}$ loop correction can be significant.
Thus the main constraint for this model comes from
neutral $K$ meson system.
%
The matrix element for the scalar operator is given by
\begin{equation}
\langle
K^{0}|\bar{d}(1-\gamma_{5})s\otimes\bar{d}(1+\gamma_{5})s|\bar{K}^{0}\rangle
=\left[\frac{1}{3}+\frac{2 m_{K}^{2}}{(m_{s}+m_{d})^{2}}\right]
f_{K}^{2}m^2_{K}B_{K}^{S}
\end{equation}
with $B^S_K$ the bag factor for scalar operator. 
%
%

Since the $c-$quark mass dominates over the $s,d$ quarks, only the
first term in eq.(18) is considered which involves left-handed CKM
matrix $V^{L}$. The dominant contributions are
\begin{eqnarray}
 H_{eff}^{H^\pm W_1+H^\pm H^\pm}
&=&\frac{G_{F}^{2}}{16\pi^{2}}m_{W}^{2} x_c y_c
(\lambda_{c}^{LL})^{2} \left[
2\eta_{cc}^{HW}|\xi_{c}|^{2}B_{V}^{HW}(y_{c},y_{W})
+\frac{1}{4}\eta_{cc}^{HH}|\xi_{c}|^{4}B_{V}^{HH}(y_{c},y_{W})
\right] \nonumber \\
& & \cdot\bar{d}\gamma^{\mu}(1-\gamma_{5})s
\otimes\bar{d}\gamma_{\mu}(1-\gamma_{5})s+\mbox{H.c} ,
\end{eqnarray}
with $y_c = m_c^2/m^2_{H^{\pm}}$ and $y_W = m_W^2/m^2_{H^{\pm}}$. The loop functions
are given by \cite{Wise80}
\begin{eqnarray}
B_{V}^{HW}(y, y_{W})&=& \frac{y_W-{1\over4}}{(1-y)(y-y_W)}
+\frac{y_W-{1\over4} y}{(1-y)^2(1-y_W)}\ln y +{3\over4} \frac{y_W^2
\ln (y_W/y)}{(y_W-y)^2(1-y_W)} ,
\nonumber\\
B_{V}^{HH}(y, y_{W})&=& \frac{1+y}{(1-y)^2} +\frac{2y}{(1-y)^3}\ln y .
\end{eqnarray}
Note that the $H^\pm H^\pm$ loop is proportional to $|\xi_{c}^{4}|$, which
significantly enhances the charged Higgs contribution at large
$\xi_{c}$.

The contributions to the effective Hamiltonian can also arise from
the flavor changing neutral current interactions via neutral Higgs
exchange at tree level, which is denoted as $H_{eff}^{h^0}$. Summing
up all the individual contributions, the total effective Hamiltonian
is
\begin{equation}
H_{eff}=H_{eff}^{W_{1}W_{1}}+H_{eff}^{W_{1}W_{2}+SW_{2}}+H_{eff}^{H^\pm W_{1}+H^\pm H^\pm}
+ H_{eff}^{h^0} .
\end{equation}

\subsection{$K^{0}-\bar{K}^{0}$ mass difference}

\begin{figure}

\includegraphics[width=0.75\textwidth]{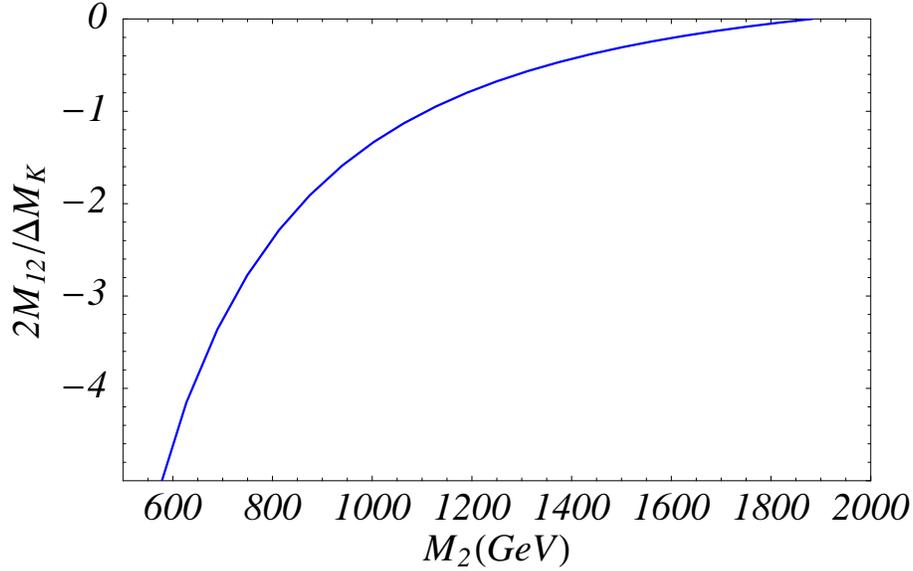}

\caption{Contribution to $2Re(M_{12})$ from right-handed $W_{R}$,
normalized to the experimental data of $\Delta m_{K}$}
\label{dmk_W2}
\end{figure}

\begin{figure}

\includegraphics[width=0.75\textwidth]{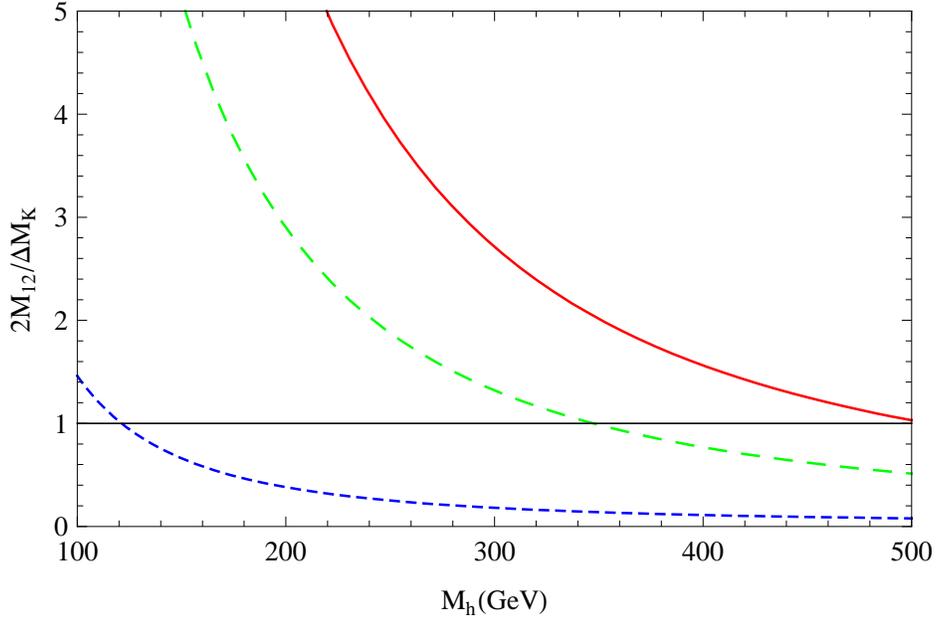}

\caption{Contribution to $2\mbox{Re}(M_{12})$ from the lightest charge
Higgs $H^{+}$. Three curves corresponds to $\xi_{c}=$30(solid), 25(
dashed) and 15 ( dotted )}. \label{dmk_H+}
\end{figure}

The mass difference for $K$ meson is simply given by
\begin{equation}
\Delta m_{K}\simeq2 \mbox{Re}(M_{12}) ,
\end{equation}
which is dominated by internal $c-$ quark loop for all loop
diagrams, and can be calculated numerically. In Fig.\ref{dmk_W2}
the individual contribution from $W_1 W_2$ diagram is shown,  the
figure indicates a large negative $M_{12}$ relative to the observed
$K^0$ mass difference $\Delta m_K$ at low $M_2<1$TeV, which reflects
the difficulties to have a light $W_2$ bellow TeV in the one Higgs
bi-doublet LR model.
The situation is different when the charged Higgs contribution is
taken into account. For the same set of CKM matrix element, the
charged Higgs gives a positive contribution to $M_{12}$, which is
comparable to the $W_2$ term for light charged Higgs mass with
sufficiently large couplings. Taking only the dominant $H^\pm H^\pm$
loop contribution, we find that a cancellation between $W_1W_2$ and
$H^\pm H^\pm $ loop requires
\begin{eqnarray}
  \eta^H_{cc} x_c |\xi_c|^4\frac{M_2^2}{m_H^2}
  &\simeq&-24
  \left[ \eta^{LR}_1 (\ln x_c+1)
    +\frac{\eta^{LR}_2}{4}\ln\frac{m_W^2}{M_2^2}
  \right]
  \frac{m_K^2}{(m_s+m_d)^2} \frac{B^S_K}{B_K} .
\end{eqnarray}
A numerical calculation including all the contributions  is shown in
Fig.\ref{dmk_H+}. Numerically, for $m_{H^+}\sim 150$GeV and Yukawa
coupling $\xi\sim 25$, the charged Higgs can compensate a opposite
contribution from a light  $W_R$ at $M_2\sim 600$ GeV. The large
Higgs contribution relies on the fact that the $H^\pm W_1$ loop is
proportional to $|\xi_c|^2$ and $H^\pm H^\pm$ loop proportional to
$|\xi_c|^4$, which grow rapidly with $|\xi^c|$ increasing.

\begin{figure}

\includegraphics[width=0.75\textwidth]{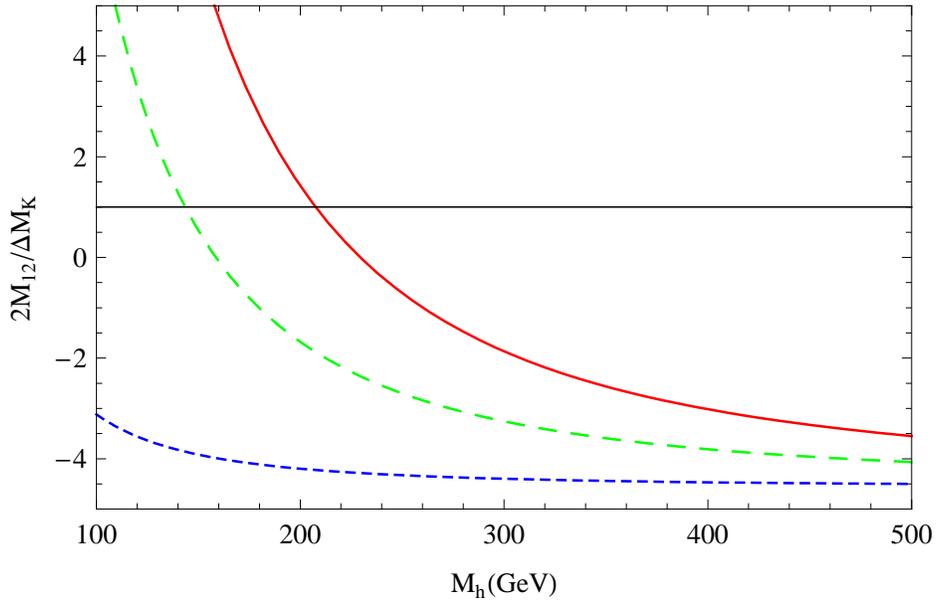}

\caption{Sum of all loop contributions, including the SM
contribution to the $2M_{12}$ normalized to $\Delta m_{K}$ with
$M_{2}=600$ GeV. Three curves corresponds to $|\xi_{c}|=$30(solid),
25( dashed) and 15 ( dotted ) respectively.}
\label{dmk_sum}
\end{figure}

\begin{figure}

\includegraphics[width=0.75\textwidth]{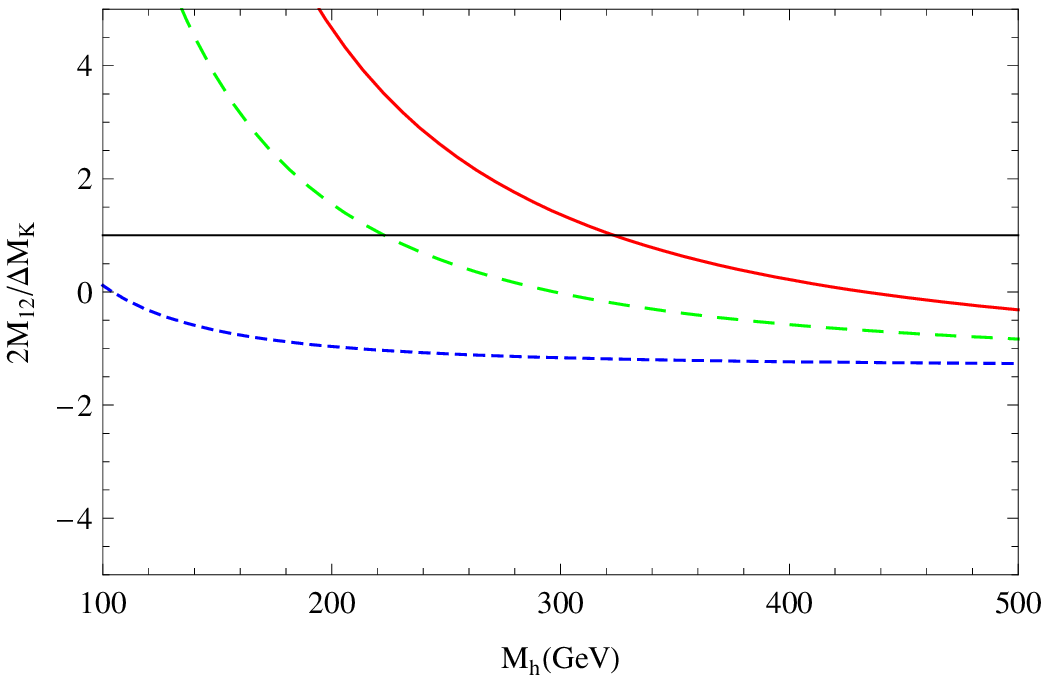}

\caption{The same as in Fig.\ref{dmk_sum} with  $M_{2}=1$TeV.}
\label{dmk_sum_1TeV}
\end{figure}

In Fig.\ref{dmk_sum} we give total loop contributions from $W_1
W_2$ loop, $W_1 H$ loop, $H H$ loop together with the $W_1 W_1$ loop
in the SM. The mass of $W_2$ is set to $600$GeV. One sees that in
the range of 150GeV$<m_H^+<250$GeV and $25<\xi_c<30$, the whole
contribution can coincide with the experimental data of $\Delta
m_K=(3.483\pm0.006)\times 10^{-15}$GeV.
%
Generically, the needed charged Higgs mass $m_{H}^+$ grows with the
mass of $W_2$. For a heavier $W_2$ at 1TeV and the same Yukawa coupling $\xi_c$
the allowed value of  $m_{H}^+$ is around $250 \sim 350$ GeV.  The numerical result
for this case is shown in Fig.\ref{dmk_sum_1TeV}.

We now check the contributions from the flavor changing neutral
current interactions via neutral Higgs exchanges at tree level. From
eq.(\ref{fcnc}), we have
\begin{equation}
M_{12}^{h^0}\simeq 6.0\times 10^{-16}
\left(
  \frac{200\mbox{GeV}}{m_{h^0}}
\right)^2
\frac{(\eta^d_{12}-\eta^{d*}_{21})^2}{(0.1)^2} ,
\end{equation}
where we have taken $m_d=9$MeV and $m_s=180$MeV. By requiring that
the $h^0$ contribution can not excess the experimental data of
$\Delta m_K$, one arrives
at a upper bound of
\begin{equation}
\frac{\sqrt{\mbox{Re}[(\eta^d_{12}-\eta^{d*}_{21})^2 ]}}{m_{h^0}}
\leq 8.3 \times 10^{-4} \mbox{GeV}^{-1} .
\end{equation}
For  $m_{h^0}$ around 200 GeV,
$\sqrt{\mbox{Re}[(\eta^d_{12}-\eta^{d*}_{21})^2 ]}\leq 0.16\approx \mathcal{O}(0.1)$,
which is in agreement with the approximate global U(1) family symmetry
of $|\eta^d_{12}|\ll 1$.  It will be shown below
that a more stringent constraint can arise from the indirect CP
violation $\epsilon_K$. From that constraint, the contributions to
the mass different from the flavor changing neutral Higgs
interactions can be neglected.

\subsection{Indirect CP violation $\varepsilon_{K}$}

The indirect CP violation parameter $\epsilon_{K}$ arises from
imaginary part of the mixing amplitude, dominated by internal
$(c,t)$ quarks and $(t,t)$ quarks. With different CP phases, the
interference among $W_1 W_1$, $W_1 W_2$ and $H^\pm H^\pm$ are rather
complicated and the allowed parameter space is large.

The expression for indirect CP violation is given by
\begin{equation}
\left|\epsilon_{K}\right|\simeq\frac{1}{2\sqrt{2}}\left(\frac{\mbox{Im}M_{12}}{\mbox{Re} M_{12}}+2\xi_{0}\right)
\simeq\frac{\mbox{Im}M_{12}}{\sqrt{2}\Delta m_{K}} ,
\end{equation}
where $\xi_0$ is the weak phase of $K\to \pi\pi$ decay amplitude
with isospin zero.

Let us first examine  the simplest case in which $|\xi_t|$ is tiny so that
$H^\pm H^\pm$ loop is negligible. In this case  the only extra contribution
is from the $W_1 W_2$ loop. It is straight forward to see that once
the $(c,c)$ loop is set real as in Eq.(\ref{phase}), the $(c,t)$ and $(t,t)$ loop
contributions become real as well because
\begin{eqnarray}
\mbox{Im}[\lambda^{LR}_c \lambda^{RL}_t +(L\to R)] &=&
-2|V^L_{cs}V^L_{cd}V^{L*}_{ts}V^{L*}_{td}|
\cos(-\alpha_2+\alpha_3-\beta_2+\phi)
\nonumber\\
&& \times \sin(\alpha_1-\alpha_2-\beta_1) \ ,
\end{eqnarray}
and
\begin{eqnarray}
\mbox{Im}(\lambda^{LR}_t\lambda^{RL}_t)&=& -|V^L_{ts}V^L_{td}|^2
\sin(\alpha_1-\alpha_2-\beta_1) \ .
\end{eqnarray}
where $\phi$ stands for
$\mbox{arg}(V^L_{cs}V^L_{cd}V^{L*}_{ts}V^{L*}_{td})$ and $\phi\simeq
\beta'_L\simeq \beta_L$ in the Wolfenstein parametrization. Thus,
there is no contribution to $\epsilon_K$ from $W_1W_2$ loop.

In the case of non-negligible $\xi_t$, the charged Higgs contributes
which involve the term proportional to $|\xi_t|^4$ and
$|\xi_t\xi_c|^2$. The constraints are very similar to the general
2HDM case which has been analyzed in details in Ref.\cite{WZ1}. For
a light charged Higgs mass $m_{H^+}=150\sim 300$ GeV, the typically
allowed values are
\begin{equation}
|\xi_c|\sim \mathcal{O}(10) \quad \mbox{and} \quad
|\xi_t|\sim \mathcal{O}(10^{-1}) ,
\end{equation}
which is consistent with the previous discussion. As $\xi_t$ is a free
parameter, the constraints on $\xi_c$ is not severe. Nevertheless, it indicates
that a small $|\xi_t|$ is needed to meeting the experiments.

The $\mbox{Im} M_{12}^{h^0}$ contribution to the indirect CP violation
$\epsilon_K$ from the flavor changing neutral current via neutral
Higgs change at tree level can be significant
\begin{equation}
 \epsilon^{h^0}_K \simeq 4.25\times 10^{-4}
\left(
  \frac{200\mbox{GeV}}{m_{h^0}}
\right)^2
\frac{Im[(\eta^d_{12}-\eta^{d*}_{21})^2]}{(0.01)^2} .
\end{equation}
From the requirement $|\epsilon_K^{h^0}| < \epsilon_K^{exp}$, we
arrive at the constraint
\begin{equation}
\frac{|\mbox{Im}\left(\eta_{12}^d - \eta_{21}^{d
\ast}\right)^{2}|^{1/2} }{ m_{h^{0}}} < 6.9\times 10^{-5}
\mbox{GeV}^{-1} ,
\end{equation}
which is a more stringent constraint on the imaginary part of the
Yukawa couplings. For $m_{h^0}=200$GeV, one has
$|\mbox{Im}\left(\eta_{12}^d - \eta_{21}^{d
\ast}\right)^{2}|^{1/2}<0.014$. It  requires that either the
off-diagonal coupling $\eta^d_{12}$ should be very small
 or CP-violating phase
must be  tuned to be very small, or the neutral scalars must be very heavy,
above 1 TeV.  Here we shall consider the small off diagonal coupling via the
mechanism of approximate U(1) family symmetry, which allows to have a
light Higgs boson at the electroweak scale. In other words, the flavor
changing neutral Higgs interactions in the two Higgs
bi-doublet model can really be suppressed via such a mechanism,
namely
\begin{equation}
 |\eta_{12}^d| \leq 0.01 , \quad \mbox{for} \quad m_{h^0}
\sim 200\ \mbox{GeV},
\end{equation}
which is significantly different from the case in the one Higgs
bi-doublet model\cite{FCNC,NSCP1,LRB6,LRB7,LRB8} in which the off
diagonal coupling is fixed  to $(V^{L\dagger})_{2i}m^d_i
(V^R)_{i1}v_1^*/(|v_1|^2-|v_2|^2)$.

\section{Neutral $B$ meson system}

In the previous section, we have illustrated that a light right-handed
gauge boson can coincide with the $K$ mixing data. In this section, we
shall show that it is consistent with the $B$ mixing measurements as
well.
Unlike the case in the Kaon system, the $B$ meson mixing is dominated
by internal $t-$quark loop. Due to the weaker dependence of loop
functions $I_{1,2}(x,\beta)$ on quark masses. $I_1(x_t,\beta)$ and
$I_2(x_t,\beta)$ from the $W_1 W_2$ loop are only a few percent of
$S_0(x_t)$, which greatly suppresses it's phenomenological
significance in $B$ mixing and decays.
The charged Higgs contribution is dominated by en extra parameter,
the Yukawa coupling $\xi_t$, and is suppressed if $|\xi_t|$ is small.
%

In the first step let us take a close look at the $W_1 W_2$
contribution.  The effective Hamiltonian for $\Delta B=2$ process with
$W_1W_2$ loop is similar to the $\Delta S=2$ case
\begin{eqnarray}
H_{eff} & \simeq & \frac{G_{F}^{2}m_{W}^{2}}{8\pi^{2}}\lambda_{t}^{LR}\lambda_{t}^{RL}\beta x_{t}
\left[(4+x_{t}^{2}\beta)I_{1}\eta_1 (x_{t},\beta)
-(1+\beta)I_{2}\eta_2 (x_{t},\beta)\right]
\nonumber\\
&  & \bar{d}(1-\gamma_{5})b\otimes\bar{d}(1+\gamma_{5})b + \mbox{H.c} .
\end{eqnarray}
The QCD corrections at scale $m_b$ are $\eta_1\simeq 1.8$ and
$\eta_2\simeq 1.7$\cite{LRB7}. The matrix element is given by
\begin{eqnarray}
M_{12}^{W_{1}W_{2}}
& = & \frac{G_{F}^{2}m_{W}^{2}}{8\pi^{2}}
\lambda_{t}^{LR}\lambda_{t}^{RL}\beta x_t
\left[4\eta_{1} I_{1}(x_{t},\beta)-\eta_{2} I_{2}(x_{t},\beta)\right]
\left(\frac{m_{B}^{2}}{m_{b}^{2}}+\frac{1}{6}\right)f_{B}^{2}m_{B}B_{B}^{S}
\nonumber\\
& = & \frac{G_{F}^{2}m_{W}^{2}}{8\pi^{2}}\left|V^L_{td}\right|^2
e^{-i(\beta_{L}+\beta_{R})}
\frac{m_t^2}{M_2^2}  \left[4\eta_{1} I_{1}(x_{t},\frac{m_{W}^{2}}{M_{2}^{2}})
 -\eta_{2} I_{2}(x_{t},\frac{m_{W}^{2}}{M_{2}^{2}})\right]\\
 &  & \times\left(\frac{m_{B}^{2}}{m_{b}^{2}}+\frac{1}{6}\right)f_{B}^{2}m_{B}B_{B}^{S} .
\end{eqnarray}
In the limit that all $\alpha$ s are vanishing, one has
$\beta_R=-\beta_L$. The bag parameters from QCD sum rule gives
$B_{B}^{S}(m_{b})/B_{B}(m_{b})=1.2\pm0.2$\cite{LRM7}. and
$f_{B}\sqrt{B_{B}}=0.228\pm0.030\pm0.010$ GeV.
%
%
The total contribution is given by
\begin{eqnarray}
M_{12} & = & M_{12}^{SM}+M_{12}^{W_{1}W_{2}}=
\frac{G_{F}^{2}m_{W}^{2}}{6\pi^{2}}\left|V^L_{td}\right|^2
e^{-2i\beta_{L}}f_{B}^{2}m_{B}B_{B} \{\eta_{B}S_{0}(x_{t})
\nonumber\\
&  &
+\frac{3}{4}e^{-i(\beta_{L}-\beta_{R})}\frac{m_{t}^{2}}{M_{2}^{2}}
\left[4\eta_{1} I_{1}(x_{t},\frac{m_{W}^{2}}{M_{2}^{2}})-\eta_{2} I_{2}(x_{t},\frac{m_{W}^{2}}{M_{2}^{2}})\right]
\left(\frac{m_{B}^{2}}{m_{b}^{2}}+\frac{1}{6}\right)\frac{B_{B}^{S}}{B_{B}}
\}
\end{eqnarray}
with QCD correction factor $\eta_B=0.551\pm0.007$.
%
The neutral $B$ meson mass difference is given by
\begin{equation}
\Delta m_{B}\simeq2\left|M_{12}\right| .
\end{equation}
The latest data give $\Delta m_{B}=(3.337\pm0.003)\times10^{-13}$ GeV.
The $B^0$ and $B^0_s$ mixing are the most important for determining
the CKM matrix elements $V_{td}$ and $V_{ts}$ in the SM.  The SM
global fit gives
$|V^L_{td}|=(7.4\pm0.8)\times10^{-3}$\cite{CKMfitter}.  In the
presence of new physics. The connection between $\Delta m_B$ and the
CKM matrix elements is in general complicated (see. e.g,\cite{CKMfitter}).
The right-handed gauge
boson will contribute to both mixing amplitude and phases.
%
%

With the pollution from $W_{2}$, the time-dependent decay $B\to J/\psi
K_{S}$ only measures an effective phase angle which may differ from
$\beta_{L}$. The expression for $\beta_{eff}$ is

\begin{equation}
2\beta_{eff}=\mbox{Im}\left(\frac{q}{p}\frac{\bar{A}}{A}\right)
=\mbox{Im}\sqrt{\frac{M_{12}^{*}}{M_{12}}}
=\mbox{arg}\left(M_{12}^{*}\right)
\end{equation}
Using the measured experimental value of $\Delta m_{B}$ and
$\beta_{eff}$ one can obtain the value of $\beta_{L}$ as a function
of $\beta_{R}$ only. In the limit
$m_{W}^{2}\ll M_{2}$, $\beta_{L}$ is close to $\beta_{eff}$ we have
in a good approximation
\begin{equation}
\tan2\beta_{L}\simeq
\tan2\beta_{eff}\left[1-r\frac{\sin(\beta_{R}-\beta_{eff})}{2\sin4\beta_{eff}}\right] ,
\end{equation}
where $r$ is the ratio between $W_{1}W_{1}$ and $W_{1}W_{2}$ box
diagrams
\begin{eqnarray}
r =
\frac{3m_{t}^{2}(4\eta_{1} I_{1}(x_{t},\beta)-\eta_{2} I_{2}(x_{t},\beta))}{4M_{2}^{2}S(x_{t})\eta_{B}}
\left(\frac{m_{B}^{2}}{m_{b}^{2}}+\frac{1}{6}\right)
\frac{B_{B}^{S}}{B_B} .
\end{eqnarray}
The above express also lead to a bound on $\beta_{L}$
expressions
\begin{equation}
\tan2\beta_{eff}\left[1-\frac{r}{2\sin4\beta_{eff}}\right] \leq \tan2\beta_{L} \leq \tan2\beta_{eff}
\left[1+\frac{r}{2\sin4\beta_{eff}}\right] .
\end{equation}

\begin{figure}

\includegraphics[width=0.7\textwidth]{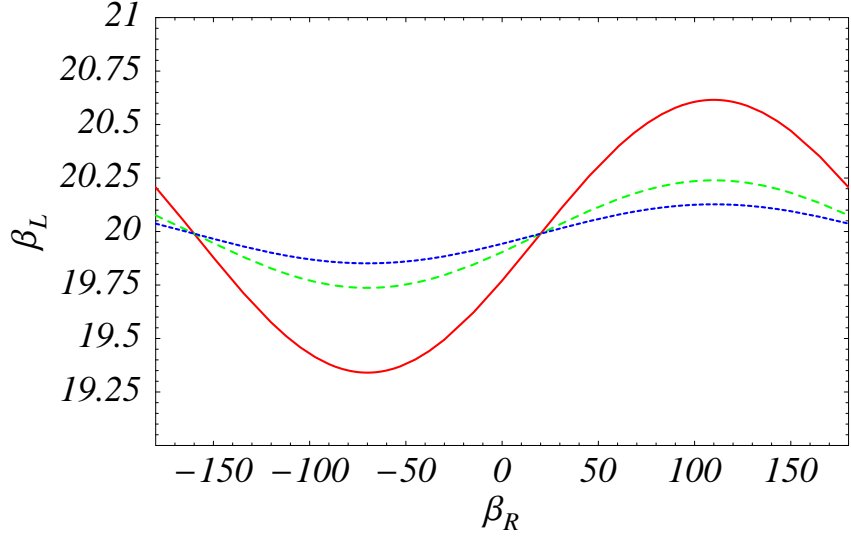}

\caption{Values of $\beta_{L}$ as a function of $\beta_{R}$ for
different $M_{2}$. Three curves correspond to $M_{2}=500$ GeV
(solid), 1000 GeV (dashed) and 1500 GeV(dotted) respectively.}
\label{betaL}
\end{figure}

\begin{figure}

\includegraphics[width=0.7\textwidth]{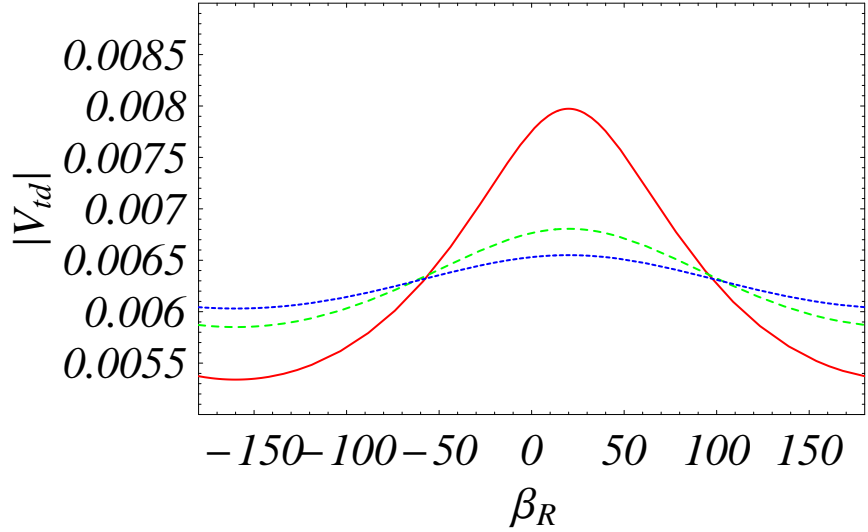}

\caption{Value of $\left|V_{td}\right|$as function of $\beta_{R}$.
Three curves correspond to $M_{2}=500$ GeV (solid), 1000 GeV
(dashed) and 1500 GeV(dotted) respectively.} \label{vtd}
\end{figure}

In Fig.(\ref{betaL}), we plot the $\beta_L$ as a function of $\beta_R$
with different right-handed gauge boson mass $M_2$. As mentioned
before, the $W_2$ contribution to $B$ mixing is rather limited. One
sees that for a light $W_2$ around 600 GeV, and $\beta_R$ varying from
$-180^\circ$ to $180^\circ$, the modification to $\beta_L$ is less
than $2^\circ$.

Once the $\beta_L$ is obtained, one can evaluate the matrix element
$|V^L_{td}|$ which given in Fig.(\ref{vtd}) as a function of
$\beta_R$.  One sees again that the changes in $|V^L_{td}|$ is small
for the whole range of $\beta_R$. Comparing with the global SM fit value
of $|V^L_{td}|$ the modifications is
within the $1\sigma$ error range.
Thus the model can easily accommodate both the data of $\Delta m_B$
and $\sin2\beta_{J/\psi}$. For the left-right model with only one Higgs
bi-doublet, since both $\beta_L$ and $\beta_R$ are calculable quantities
which depends only on the quark masses and ratios of VEVs, there is
little room to meet the CP violation in both $K$ and $B$ system. Due to
the suppression of CP phase form $\epsilon_K$, the predicted
$\sin2\beta_{J/\psi}$ has to be small and can not excess $0.1$.

%
%
The constraints from $B$ system to charged Higgs couplings is quite
similar to the general 2HDM. For $\xi\ll 1$, the box-diagram contribution
can be safely neglected.
%
%
The constraint on the neutral Higgs FCNC couplings can be easily
obtained from Eq.(\ref{fcnc}).
\begin{equation}
M_{12}^{h^0}\simeq 1.0\times 10^{-12}
\left(
  \frac{200\mbox{GeV}}{m_{h^0}}
\right)^2
(\eta^d_{13}-\eta^{d*}_{31})^2.
\end{equation}
Using the experimental data $\Delta m_B=3.337\times 10^{-3}$, one get
a upper bound of
\begin{equation}
\frac{|\eta^d_{13}-\eta^{d*}_{31}|}{m_{h^0}}
\leq 2.4 \times 10^{-3} \mbox{GeV}^{-1} .
\end{equation}
For typical $m_{h^0}=200$GeV,
$|\eta^d_{13}-\eta^{d*}_{31}|<0.41$.

Similarly, from the recently measured
$\Delta m_{B_s}=17.77\pm 0.01\pm0.07\mbox{ps}^{-1}$\cite{Evans:2007hq}, one can infer a bound for
the couplings $\eta^d_{23}$ and $\eta^d_{32}$
\begin{equation}
\frac{|\eta^d_{23}-\eta^{d*}_{32}|}{m_{h^0}}
\leq 2.7 \times 10^{-3} \mbox{GeV}^{-1} .
\end{equation}
For typical $m_{h^0}=200$GeV, we get
$|\eta^d_{13}-\eta^{d*}_{31}|<0.54$. Both bounds satisfies the
condition of $|\eta^q_{ij}|\ll 1$ from the approximate global $U(1)$
family symmetry.

\section{Conclusions}

In summary, motivated by natural spontaneous P and CP violation and
the latest low energy experimental results, we have investigated a
general left-right symmetric model with two Higgs bi-doublets. This
simple extension evades the stringent constraints from $K$ meson
mixing, and lowers the allowed mass of right-handed gauge boson
closing to the current direct experimental search bound $\sim 600$GeV.
Through a negative interference with charged Higgs loop, which
automatically occur when the charged Higgs is also light around
electroweak scale with large Yukawa couplings. The FCNC can be suppressed by the mechanism of
approximate global $U(1)$ family symmetry. We have illustrated that
the off diagonal Yukawa couplings of $\mathcal{O}(10^{-2})\sim
\mathcal{O}(10^{-1})$ are consistent with all the constraints.  This
model has rich sources of CP violation which may show up in lower
energy processes such as rare $B$ decays and the new physics particles
can be directly searched in upcoming LHC and future ILC experiments.

\acknowledgments

\label{ACK}

We would like to thank T.D. Lee for suggesting the study of
spontaneous P and CP violations. We are grateful to Y.-P. Kuang for
valuable discussions and suggestions. This work is supported in part
by the National Science Foundation of China (NSFC) under the grant
10475105, 10491306, and the key Project of Chinese Academy of
Sciences (CAS). YFZ is supported by JSPS foundation.


\begin{thebibliography}{99}
\bibitem{LY}
  T.~D.~Lee and C.~N.~Yang,
  ``Question Of Parity Conservation In Weak Interactions,''
  Phys.\ Rev.\  {\bf 104}, 254 (1956).

\bibitem{CSW}
  C.~S.~Wu, E.~Ambler, R.~W.~Hayward, D.~D.~Hoppes and R.~P.~Hudson,
  ``EXPERIMENTAL TEST OF PARITY CONSERVATION IN BETA DECAY,''
  Phys.\ Rev.\  {\bf 105}, 1413 (1957).


\bibitem{CP}
%
  J.~H.~Christenson, J.~W.~Cronin, V.~L.~Fitch and R.~Turlay,
  ``Evidence For The 2 Pi Decay Of The K(2)0 Meson,''
  Phys.\ Rev.\ Lett.\  {\bf 13}, 138 (1964).


\bibitem{SM-1}
  S.~L.~Glashow,
  ``Partial Symmetries Of Weak Interactions,''
  Nucl.\ Phys.\  {\bf 22}, 579 (1961).


\bibitem{SM-2}
  S.~Weinberg,
  ``A Model Of Leptons,''
  Phys.\ Rev.\ Lett.\  {\bf 19}, 1264 (1967).

\bibitem{SM-3}
A. Salam, in {\it Proceedings of the Eight Nobel Symposium}, edited
by N. Svartholm (Almqvist and Wikell, Stockholm, 1968).


\bibitem{KM}
  M.~Kobayashi and T.~Maskawa,
  ``CP Violation In The Renormalizable Theory Of Weak Interaction,''
  Prog.\ Theor.\ Phys.\  {\bf 49}, 652 (1973).


\bibitem{YLW1-1}
  Y.~L.~Wu,
  ``A new prediction for direct CP violation epsilon'/epsilon and Delta(I)  =
  1/2 rule,''
  Phys.\ Rev.\  D {\bf 64}, 016001 (2001)
  [arXiv:hep-ph/0012371].

\bibitem{YLW1-2}
For a review, see: S. Bertolini, "Theory Status
of $\varepsilon'/\varepsilon$ ", Frascati Phys. Ser. 28 275-290
ги2002), hep-ph/0206095.


\bibitem{NA48}
%
  J.~R.~Batley {\it et al.}  [NA48 Collaboration],
  ``A precision measurement of direct CP violation in the decay of neutral
  kaons into two pions,''
  Phys.\ Lett.\  B {\bf 544}, 97 (2002)
  [arXiv:hep-ex/0208009].

\bibitem{KTeV}
  A.~Alavi-Harati {\it et al.}  [KTeV Collaboration],
  ``Measurements of direct CP violation, CPT symmetry, and other parameters in
  the neutral kaon system,''
  Phys.\ Rev.\  D {\bf 67}, 012005 (2003)
  [Erratum-ibid.\  D {\bf 70}, 079904 (2004)]
  [arXiv:hep-ex/0208007].


\bibitem{WZZ1}
Y.~L.~Wu and Y.~F.~Zhou,
``Implications of charmless B decays with large direct CP violation,''
Phys.\ Rev.\  D {\bf 71}, 021701 (2005);
%
\bibitem{WZZ2}Y.~L.~Wu and Y.~F.~Zhou,
``Charmless decays B $\to$ pi pi, pi K and K K in broken SU(3)
symmetry,'' Phys.\ Rev.\  D {\bf 72}, 034037 (2005);

\bibitem{WZZ3}Y.~L.~Wu, Y.~F.~Zhou and C.~Zhuang,
``Implications of new data in charmless B decays,'' Phys.\ Rev.\  D
{\bf 74}, 094007 (2006).



\bibitem{BB1}
K.~Abe {\it et al.}
  K.~Abe {\it et al.}  [Belle Collaboration],
  ``Observation of large CP violation and evidence for direct CP violation  in
  B0 $\to$ pi+ pi- decays,''
  Phys.\ Rev.\ Lett.\  {\bf 93}, 021601 (2004)
  [arXiv:hep-ex/0401029].



\bibitem{BB2}  B.~Aubert {\it et al.}  [BaBar Collaboration],
  ``Observation of direct CP violation in $B^0 \to K^+ \pi^-$ decays,''
  Phys.\ Rev.\ Lett.\  {\bf 93}, 131801 (2004)
  [arXiv:hep-ex/0407057].



\bibitem{TDL-1}
  T.~D.~Lee,
  ``A Theory of Spontaneous T Violation,''
  Phys.\ Rev.\  D {\bf 8}, 1226 (1973).
\bibitem{TDL-2}
  T.~D.~Lee,
  ``CP Nonconservation and Spontaneous Symmetry Breaking,''
  Phys.\ Rept.\  {\bf 9}, 143 (1974).




\bibitem{3HDM}
  S.~Weinberg,
  Phys.\ Rev.\ Lett.\  {\bf 37}, 657 (1976).



\bibitem{HW}
  L.~J.~Hall and S.~Weinberg,
  ``Flavor changing scalar interactions,''
  Phys.\ Rev.\  D {\bf 48}, 979 (1993)
  [arXiv:hep-ph/9303241].


\bibitem{WW}
  Y.~L.~Wu and L.~Wolfenstein,
  ``Sources of CP violation in the two Higgs doublet model,''
  Phys.\ Rev.\ Lett.\  {\bf 73}, 1762 (1994)
  [arXiv:hep-ph/9409421].


\bibitem{WU} Y.L. Wu, Carnegie-Mellon Univ. report, CMU-HEP94-01, hep-ph/9404241, 1994;
\\ Invited talk at 5th Conference on the Intersections
of Particle and Nuclear Physics, St. Petersburg, FL, 31 May- 6 Jun
1994, published in Proceedings, pp338, edited by S.J. Seestrom (AIP,
New York, 1995), hep-ph/9406306.
\bibitem{WW1}
  L.~Wolfenstein and Y.~L.~Wu,
  ``CP violation in the decay b $\to$ s gamma in the two Higgs doublet model,''
  Phys.\ Rev.\ Lett.\  {\bf 73}, 2809 (1994)
  [arXiv:hep-ph/9410253].




\bibitem{WU1}
  Y.~L.~Wu,
  ``Probing new physics from CP violation in radiative B decays,''
  Chin.\ Phys.\ Lett.\  {\bf 16}, 339 (1999)
  [arXiv:hep-ph/9805439].




\bibitem{WZ1}

  Y.~L.~Wu and Y.~F.~Zhou,
  ``F0 anti-F0 mixing and CP violation in the general two Higgs doublet
  model,''
  Phys.\ Rev.\  D {\bf 61}, 096001 (2000)
  [arXiv:hep-ph/9906313].

\bibitem{WZ2} 
  Y.~L.~Wu and Y.~F.~Zhou,
  ``Muon anomalous magnetic moment in the standard model with two Higgs
  doublets,''
  Phys.\ Rev.\  D {\bf 64}, 115018 (2001)
  [arXiv:hep-ph/0104056].


\bibitem{WZ3} 
  Y.~L.~Wu and C.~Zhuang,
  ``Exclusive B $\to$ P V decays and CP violation in the general
  two-Higgs-doublet model,''
  Phys.\ Rev.\  D {\bf 75}, 115006 (2007)
  [arXiv:hep-ph/0701072].



\bibitem{LRM1} 
  J.~C.~Pati and A.~Salam,
  ``Lepton Number As The Fourth Color,''
  Phys.\ Rev.\  D {\bf 10}, 275 (1974)
  [Erratum-ibid.\  D {\bf 11}, 703 (1975)].

\bibitem{LRM2} 
  R.~N.~Mohapatra and J.~C.~Pati,
  ``Left-Right Gauge Symmetry And An Isoconjugate Model Of CP Violation,''
  Phys.\ Rev.\  D {\bf 11}, 566 (1975).

\bibitem{LRM3} 
  G.~Senjanovic and R.~N.~Mohapatra,
  ``Exact Left-Right Symmetry And Spontaneous Violation Of Parity,''
  Phys.\ Rev.\  D {\bf 12}, 1502 (1975).


\bibitem{LRM4} 
  D.~Chang,
  ``A Minimal Model Of Spontaneous CP Violation With The Gauge Group SU(2)-L X
  SU(2)-R X U(1)-(B-L),''
  Nucl.\ Phys.\  B {\bf 214}, 435 (1983).


\bibitem{LRM5}
  J.~M.~Frere, J.~Galand, A.~Le Yaouanc, L.~Oliver, O.~Pene and J.~C.~Raynal,
  ``K0 anti-K0 in the SU(2)-L x SU(2)-R x U(1) model of CP violation,''
  Phys.\ Rev.\  D {\bf 46}, 337 (1992).


\bibitem{LRM6} 
  G.~Barenboim, J.~Bernabeu and M.~Raidal,
  ``Spontaneous CP-violation in the left-right model and the kaon system,''
  Nucl.\ Phys.\  B {\bf 478}, 527 (1996)
  [arXiv:hep-ph/9608450].


\bibitem{LRM7}
  P.~Ball, J.~M.~Frere and J.~Matias,
  ``Anatomy of mixing-induced CP asymmetries in left-right-symmetric models
  with spontaneous CP violation,''
  Nucl.\ Phys.\  B {\bf 572}, 3 (2000)
  [arXiv:hep-ph/9910211].


\bibitem{FCNC} 
  M.~E.~Pospelov,
  ``FCNC in left-right symmetric theories and constraints on the  righthanded
  scale,''
  Phys.\ Rev.\  D {\bf 56}, 259 (1997)
  [arXiv:hep-ph/9611422].

\bibitem{NSCP1} 
  N.~G.~Deshpande, J.~F.~Gunion, B.~Kayser and F.~I.~Olness,
  ``Left-right symmetric electroweak models with triplet Higgs,''
  Phys.\ Rev.\  D {\bf 44}, 837 (1991).



\bibitem{NSCP2} 
  G.~Barenboim, M.~Gorbahn, U.~Nierste and M.~Raidal,
  ``Higgs sector of the minimal left-right symmetric model,''
  Phys.\ Rev.\  D {\bf 65}, 095003 (2002)
  [arXiv:hep-ph/0107121].


\bibitem{LRM8} 
  P.~Langacker and S.~Uma Sankar,
  ``Bounds on the Mass of W(R) and the W(L)-W(R) Mixing Angle xi in General
  SU(2)-L x SU(2)-R x U(1) Models,''
  Phys.\ Rev.\  D {\bf 40}, 1569 (1989).

\bibitem{LRM9}
  G.~Barenboim, J.~Bernabeu, J.~Prades and M.~Raidal,
  ``Constraints on the W(R) mass and CP-violation in left-right models,''
  Phys.\ Rev.\  D {\bf 55}, 4213 (1997)
  [arXiv:hep-ph/9611347].


\bibitem{LRM10} 
  Y.~Zhang, H.~An, X.~Ji and R.~N.~Mohapatra,
  ``Right-handed quark mixings in minimal left-right symmetric model with
  general CP violation,''
  Phys.\ Rev.\  D {\bf 76}, 091301 (2007)
  [arXiv:0704.1662 [hep-ph]].


\bibitem{MR-1} 
  T.~D.~Lee and C.~N.~Yang,
  ``Question Of Parity Conservation In Weak Interactions,''
  Phys.\ Rev.\  {\bf 104}, 254 (1956).

\bibitem{MR-2}
 L. Kobzarev, L. Okun and I. Pomeranchuk, Sov. J. Nucl.Phys. {\bf 3} 837 (1966).


\bibitem{SO32-1} 
  Y.~L.~Wu,
  ``Maximally symmetric minimal unification model SO(32) with three  families
  in ten dimensional space-time,''
  Mod.\ Phys.\ Lett.\  A {\bf 22}, 259 (2007)
  [arXiv:hep-ph/0607336].


\bibitem{SO32-2}
  Y.~L.~Wu,
  ``Stability of proton and maximally symmetric minimal unification model  for
  basic forces and building blocks of matter,''
  Sci.\ China {\bf G50}, 303 (2007)
  [arXiv:hep-ph/0505010].


\bibitem{BL} 
  G.~C.~Branco and L.~Lavoura,
  ``Natural CP Breaking In Left-Right Symmetric Theories,''
  Phys.\ Lett.\  B {\bf 165}, 327 (1985).

\bibitem{Akeroyd:2006bb}
  A.~G.~Akeroyd, M.~Aoki and Y.~Okada,
  ``Lepton Flavour Violating tau Decays in the Left-Right Symmetric Model,''
  Phys.\ Rev.\  D {\bf 76}, 013004 (2007).


\bibitem{CS} 
  T.~P.~Cheng and M.~Sher,
  ``Mass Matrix Ansatz and Flavor Nonconservation in Models with Multiple Higgs
  Doublets,''
  Phys.\ Rev.\  D {\bf 35}, 3484 (1987).

\bibitem{Inami:1981fz} 
  T.~Inami and C.~S.~Lim,
  `E`ffects Of Superheavy Quarks And Leptons In Low-Energy Weak Processes K(L)
  $\to$ Mu Anti-Mu, K+ $\to$ Pi+ Neutrino Anti-Neutrino And K0 $\to$ Anti-K0,''
  Prog.\ Theor.\ Phys.\  {\bf 65}, 297 (1981)
  [Erratum-ibid.\  {\bf 65}, 1772 (1981)].



\bibitem{QCDC}
  S.~Herrlich and U.~Nierste,
  ``The Complete$ |\Delta S|$=2 Hamiltonian in the Next-To-Leading Order,''
  Nucl.\ Phys.\  B {\bf 476}, 27 (1996)
  [arXiv:hep-ph/9604330].


\bibitem{Bag}
  L.~Lellouch,
  ``Light hadron weak matrix elements,''
  Nucl.\ Phys.\ Proc.\ Suppl.\  {\bf 94}, 142 (2001)
  [arXiv:hep-lat/0011088].



\bibitem{LRB1}
  G.~Beall, M.~Bander and A.~Soni,
  ``Constraint On The Mass Scale Of A Left-Right Symmetric Electroweak Theory
  From The K(L) K(S) Mass Difference,''
  Phys.\ Rev.\ Lett.\  {\bf 48}, 848 (1982).


\bibitem{LRB2}
  R.~N.~Mohapatra, G.~Senjanovic and M.~D.~Tran,
  ``Strangeness Changing Processes And The Limit On The Right-Handed Gauge
  Boson Mass,''
  Phys.\ Rev.\  D {\bf 28}, 546 (1983).


\bibitem{LRB3} 
  F.~J.~Gilman and M.~H.~Reno,
  ``Restrictions From The Neutral K And B Meson Systems On Left-Right Symmetric
  Gauge Theories,''
  Phys.\ Rev.\  D {\bf 29}, 937 (1984).


\bibitem{LRB4} 
  D.~Chang, J.~Basecq, L.~F.~Li and P.~B.~Pal,
  ``Comment On The K(L) K(S) Mass Difference In Left-Right Model,''
  Phys.\ Rev.\  D {\bf 30}, 1601 (1984).

\bibitem{LRB5} 
  W.~S.~Hou and A.~Soni,
  ``Gauge Invariance Of The K(L) K(S) Mass Difference In Left-Right Symmetric
  Model,''
  Phys.\ Rev.\  D {\bf 32}, 163 (1985).


\bibitem{LRB6} 
  J.~Basecq, L.~F.~Li and P.~B.~Pal,
  ``Gauge Invariant Calculation Of The K(L) K(S) Mass Difference In The
  Left-Right Model,''
  Phys.\ Rev.\  D {\bf 32}, 175 (1985).

\bibitem{LRB7} 
  G.~Ecker and W.~Grimus,
  ``CP Violation And Left-Right Symmetry,''
  Nucl.\ Phys.\  B {\bf 258}, 328 (1985).

\bibitem{LRB8} 
  S.~Sahoo, L.~Maharana, A.~Roul and S.~Acharya,
  ``The Masses Of W(R) Triplet Higgs, And Z-Prime Bosons In The Left-Right
  Symmetric Model,''
  Int.\ J.\ Mod.\ Phys.\  A {\bf 20}, 2625 (2005).

\bibitem{Wise80}
  L.~F.~Abbott, P.~Sikivie and M.~B.~Wise,
  ``Constraints On Charged Higgs Couplings,''
  Phys.\ Rev.\  D {\bf 21}, 1393 (1980).

\bibitem{CKMfitter}
  J.~Charles {\it et al.}  [CKMfitter Group],
  ``CP violation and the CKM matrix: Assessing the impact of the asymmetric  B
  factories,''
  Eur.\ Phys.\ J.\ C {\bf 41}, 1 (2005)
  [arXiv:hep-ph/0406184].
  Updated results may be found on the web site:
  {\tt http://ckmfitter.in2p3.fr/}.

\bibitem{Evans:2007hq}
  H.~G.~Evans  [CDF Collaboration],
  ``Bs Physics at CDF and D0,''
  arXiv:0705.4598 [hep-ex].
\end{thebibliography}
\end{document}